\documentclass[9pt,twocolumn,twoside]{osajnl}
\usepackage{amsmath,amsfonts,graphicx}
\usepackage{siunitx}
\graphicspath{{./Figs/}}
\journal{ol}
\setboolean{shortarticle}{true}
\title{Nonlinear anti-directional couplers with gain and loss}
\author[1,*]{A. Govindarajan}
\author[2]{Boris A. Malomed}
\author[1]{M. Lakshmanan}
\affil[1]{Centre for Nonlinear Dynamics, School of Physics, Bharathidasan University, Tiruchirappalli - 620 024, India}
\affil[2]{Department of Physical Electronics, School of Electrical Engineering, Faculty of Engineering, and the Center for Light-Matter Interaction, Tel Aviv University, 69978 Tel Aviv, Israel}
\affil[*]{Corresponding author: govin.nld@gmail.com}
\dates{}
\ociscodes{(190.1450)   Bistability; (350.3618)   Left-handed materials; (190.7110); (160.3918)   Metamaterials; (230.4320)   Nonlinear optical devices.}
\doi{\url{}}
\begin{abstract}
Following the concept of $\mathcal{PT}$-symmetric couplers, we propose a
linearly coupled system of nonlinear waveguides, made of positive- and
negative-index materials, which carry, respectively, gain and loss. We
report novel bi- and multi-stability states pertaining to transmitted and reflective intensities, which are controlled by the ratio of the gain and loss coefficients, and phase mismatch
between the waveguides. These states offer transmission regimes with
extremely low threshold intensities for transitions between coexisting
states, and very large amplification ratio between the input and output
intensities leading to an efficient way of controlling light with light.
\end{abstract}
\setboolean{displaycopyright}{true}
\begin{document}
\maketitle
Optical systems with separated and globally balanced gain and loss,
realizing the parity-time ($\mathcal{PT}$) symmetry \cite%
{bender1,dorey,bender2,PT1,PT2}, have drawn widespread interest in
theoretical and experimental studies, as they exhibit rich phenomenology in
the context of linear and nonlinear optics, see reviews \cite%
{bender3,review,ptqm,Konotop2016RMP,PTsol2,el2018non,PhysD} and references
therein. In particular, directional couplers, alias \textit{dimers}, which
are composed of cores carrying equal amounts of gain and loss, serve as
prime objects to explore the effects of $\mathcal{PT}$-symmetry, in the
continuous-wave (CW) \cite%
{Ruter2010Nat,Kevr-dimer,Barash-dimer,Blas,Govind_OL} and soliton \cite%
{Driben,Barash} regimes. In the CW form, directional couplers (built without
gain and loss) do not manifest any bistability, even when the Kerr
nonlinearity is included \cite{Jensen1982CCR} (although it may be made
possible by additional ingredients, such as prism coupling in directional
couplers \cite{stegeman1988bistability} or saturable nonlinearity \cite%
{Snyder}).

\begin{figure}[t]
\begin{center}
\includegraphics[width=1\columnwidth]{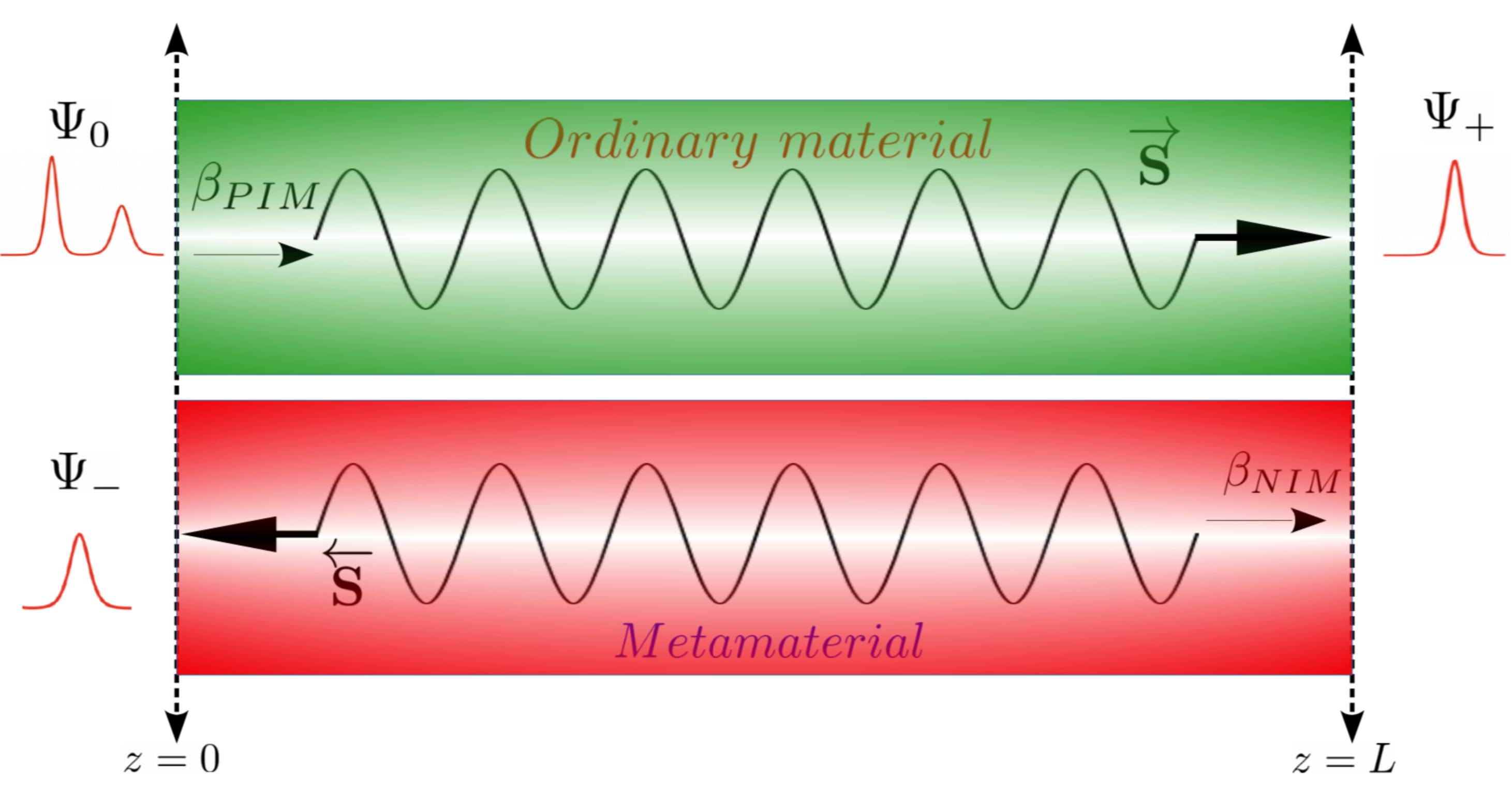}
\end{center}
\caption{A schematic of the CW propagation in the anti-directional coupler.
The green and red colors designate the PIM and NIM waveguides with intrinsic
gain and loss, respectively. Depending on its peak intensity,
the incident signal, coupled into the input 
port of the PIM waveguide, appears at the output (input) port of the
PIM (NIM) waveguide.}
\label{PIM:NIM}
\end{figure}

Directional couplers with left-handed materials, which are supposed to add
new dimensions to physics of light \cite{NL2012}, including cloaking \cite%
{Cloaking2006Sci}, magnetism at optical frequencies, reversed Snell's law,
reversed Goos-H\"{a}nchen shift, etc. \cite{Engheta2006,maimistov2007nonlinear,lapine2014}, can exhibit novel
nonlinear effects, such as optical bistability and gap solitons, the latter
usually occurring in periodic structures \cite{winfulob,Cerda} and nonlinear
Bragg gratings \cite{Aceves,Christo,deSterke,Krug}.
In contrast to the conventional couplers, which are made of two
positive-index material (PIM) waveguides, the metamaterial-including
dual-core systems may be termed\textit{\ anti-directional couplers} (ADCs),
as the propagation dynamics in a negative-index-material (NIM) waveguide is
opposite to that in conventional media. The latter fact is elucidated by the
opposite direction of the Poynting vector ($\vec{S}$ in Fig. \ref{PIM:NIM})
in NIM, which leads to the opposite phase velocity and energy flow \cite{NLG}.

As is well known, NIMs are artificial materials which exhibit inherent loss
due to the fact that their permittivity and permeability have imaginary
parts \cite{Engheta2006}.
This detrimental effect may render ADCs impractical as effective
switching devices, even if experimental verification of their operation
remains a relevant objective. Thus, an important issue is how the absorption
in NIMs affects nonlinear dynamical effects in ADC and how one can
compensate the loss, to make the device more useful for all-optical signal
processing, including ultrafast switching and memory applications. Here we
address this issue by introducing gain in the positive-index waveguide, as
motivated by the above-mentioned studies of the $\mathcal{PT}$-symmetric
couplers. The settings with both equal and different values of the gain and
loss in the two cores will be considered. As a result, we report new
dynamical regimes provided by the so designed ADCs.

We start the analysis of the ADC model with intrinsic loss ($\chi _{2}$) and
gain ($\chi _{1}$) in its NIM and PIM cores. Following Ref. \cite{NLG}
(where the gain and loss were not included), the corresponding coupled-mode
equations representing the light propagation in the system are written as
\begin{gather}
+i\frac{d \mathbf{\Psi} _{+}}{\partial z}+\gamma _{1}|\Psi _{+}|^{2}\mathbf{%
\Psi }_{+}+\kappa \mathbf{\Psi }_{-}+\left( \frac{\delta }{2}-i\chi
_{1}\right) \mathbf{\Psi }_{+}=0  \label{eqn:3} \\
-i\frac{d \mathbf{\Psi }_{-}}{\partial z}+\gamma _{2}|\mathbf{\Psi }%
_{-}|^{2}\mathbf{\Psi }_{-}+\kappa \mathbf{\Psi }_{+}+\left( \frac{\delta }{2%
}+i\chi _{2}\right) \mathbf{\Psi }_{-}=0  \label{eqn:4}
\end{gather}%
where $\mathbf{\Psi} _{+}(z)$ and $\mathbf{\Psi} _{-}(z)$ are complex envelope amplitudes of
the forward and backward waves in the PIM and NIM cores, respectively, $z$
is the propagation distance, and $\kappa $ is the inter-core linear-coupling
parameter. The phase-mismatch parameter, $\delta $, is defined by the PIM
and NIM propagation constants, $\delta =\beta _{\mathrm{PIM}}-\beta _{%
\mathrm{NIM}}$, and Kerr terms in the two cores are represented by
coefficients $\gamma _{1}$ and $\gamma _{2}$.
\begin{figure}[t]
\begin{center}
\includegraphics[width=1\columnwidth]{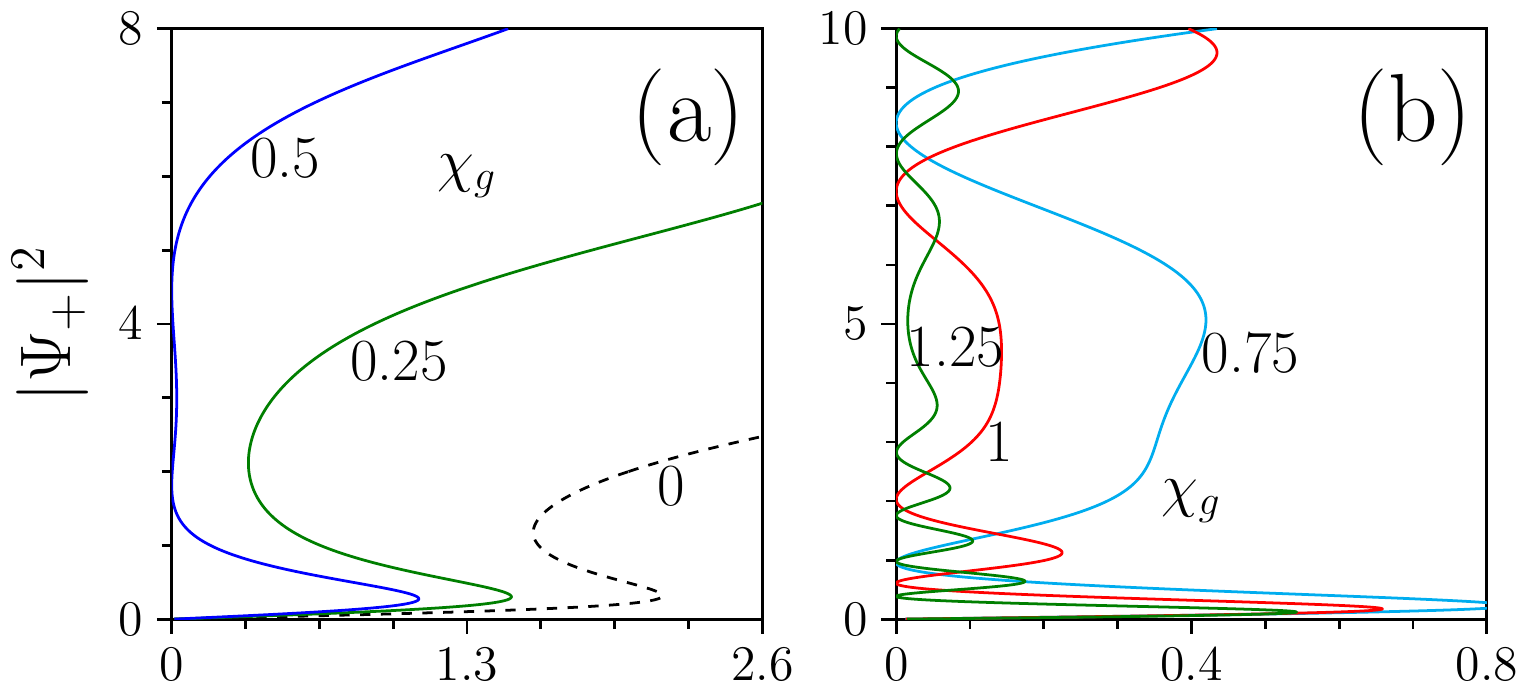} %
\includegraphics[width=1\columnwidth]{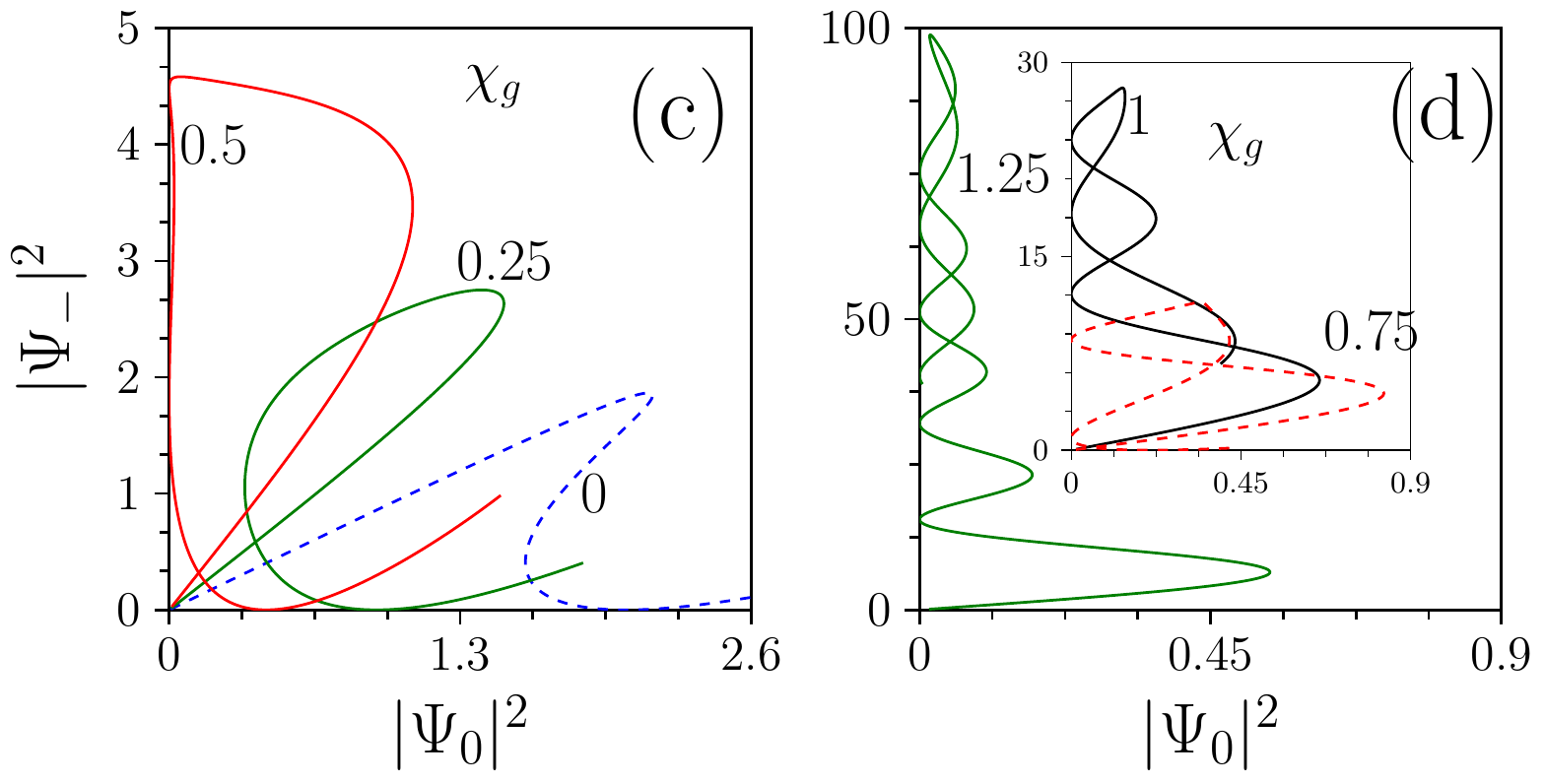}
\end{center}
\caption{Panels (a) and (b) display several bi- and multi- stable
transmissions in the system with equal gains in the two cores ($\protect\chi %
_{1}=-\protect\chi _{2}\equiv \protect\chi _{g}$), while panels (c) and (d)
portray the bi- and multi-stable reflections for the system parameters $%
\protect\kappa =1$, $\protect\delta =0$, and $\protect\gamma _{1}=\protect%
\gamma _{2}=1$ in Eqs. (\protect\ref{eqn:3}) and (\protect\ref{eqn:4}). The
total propagation length is $L=2$. Particular values of $\protect\chi _{g}$
are attached to the curves, and are also distinguished by different colors
and curve types (continuous or dashed).}
\label{figure1}
\end{figure}
Coupled-mode equations \eqref{eqn:3} and \eqref{eqn:4} were simulated using 
an implicit finite-difference method \cite{leve2007finite} with boundary conditions 
$\mathbf{\Psi}_+(0)=\mathbf{\Psi}_0$ and $\mathbf{\Psi}_-(L)=0$. The objective was to produce the transmissivity ($\Im $) 
and reflectivity ($\Re $) defined as $\Im =|%
\mathbf{\Psi }_{+}(z=L)/\mathbf{\Psi }_{0}|^{2}$ and $\Re =|\mathbf{\Psi }%
_{-}(z=0)/\mathbf{\Psi }_{0}|^{2}$, where $|\mathbf{\Psi }_{0}|^{2}\equiv
\left\vert \mathbf{\Psi }_{+}(z=0)\right\vert^2 $ is the incident
intensity of light, coupled at $z=0$ into the PIM core, and $L$ is the ADC's
length, see Fig. \ref{PIM:NIM}.
\begin{figure}[t]
\begin{center}
\includegraphics[width=1\columnwidth]{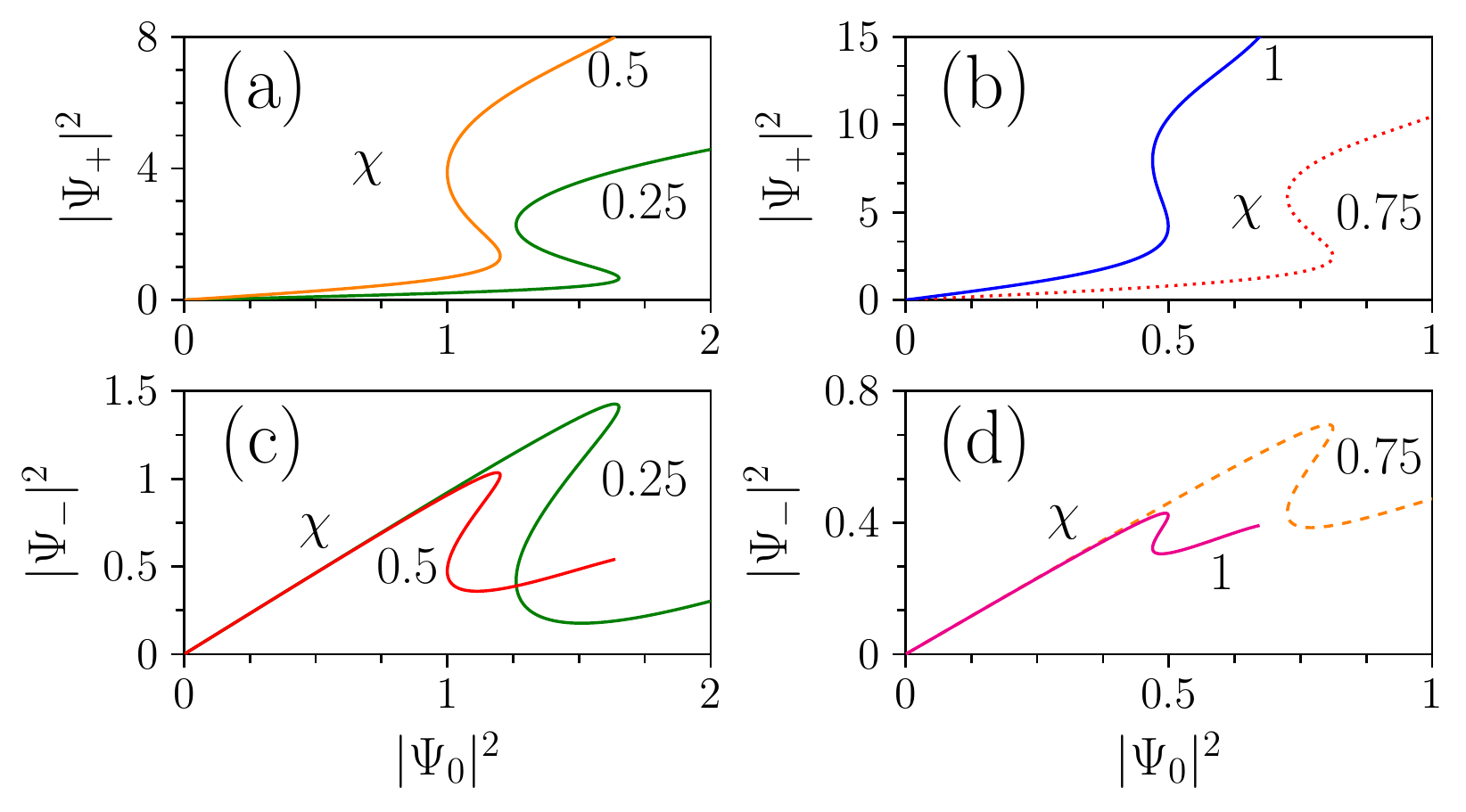}  \includegraphics[width=1%
\columnwidth]{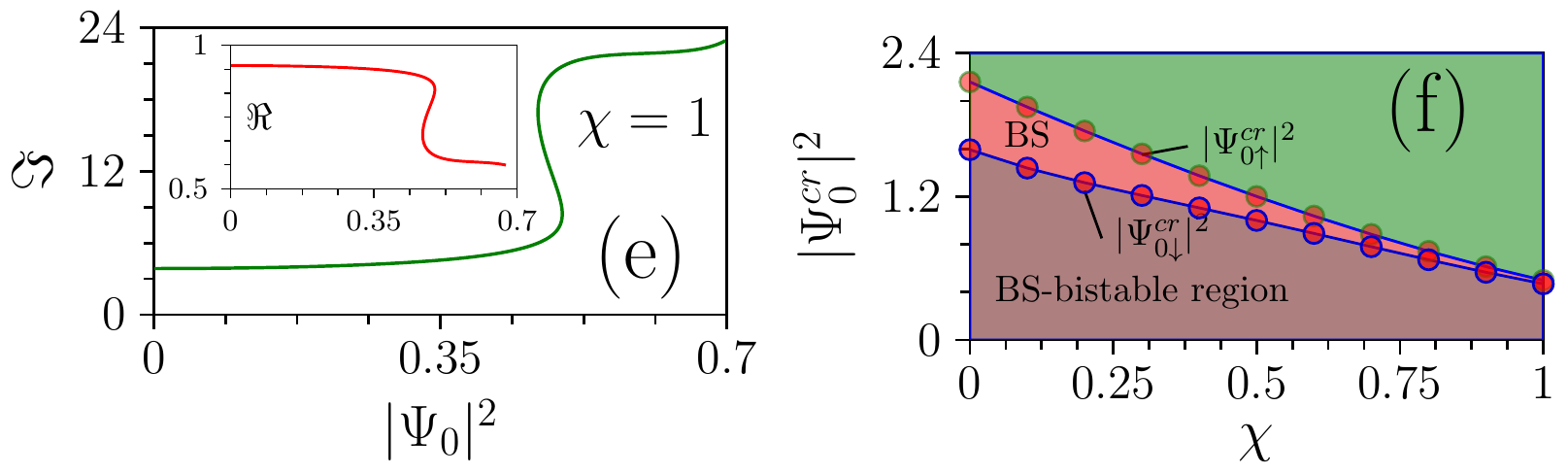}
\end{center}
\caption{Top panels (a, b) depict the bistable dynamics of transmitted wave
in the system with the balanced gain and loss, $\protect\chi _{1}=\protect%
\chi _{2}\equiv \protect\chi $, as indicated in Fig. \protect\ref{PIM:NIM}.
The dynamics representing the reflective bistability is shown in middle
panels (c, d). In the bottom panels, the plot (e) displays the
transmissivity and reflectivity curves, and (f) shows critical intensities
for the jumps down and up, as a function of $\protect\chi $, with the
bistability area between the two curves. The parameters are same as in Fig.
\protect\ref{figure1}. Particular values of $\protect\chi $ are attached to
the curves, and are distinguished by different colors and curve types.}
\label{figure2}
\end{figure}

First, in Fig. \ref{figure1} we present the results for the output intensity in
the PIM core vs. the input value, for the PIM-NIM coupler with equal gain
coefficients in both curves $\chi _{1}=-\chi _{2}\equiv \chi _{g}$.  Compared to the
optical bistability found at $\chi _{g}=0$ (in the conservative system
considered in Ref. \cite{NLG}), the bistability takes a more pronounced
shape, along with a substantial reduction in the switch-up and switch-down
threshold, even when the gain-loss coefficient takes a relatively small
value, $\chi _{g}=0.25$. This behavior, different from usual hysteresis
loops, in which one normally finds more pronounced bistable response along
with a larger switch-up threshold intensity \cite{Iooss}, persists up to $%
\chi _{g}=0.5$, where the switching characteristic features a nearly flat
vertical segment. Such a novel variety of the optical bistability may find
applications to the design of ultra-fast switching and optical memory, as
well as signal regenerators. There is a critical value, $\chi _{g}^{\mathrm{%
cr}}\approx0.5$, above which the system gives rise to new forms of multistability,
along with a dramatic drop in the lowest critical value of the intensity. When the gain-loss factor is raised to $\chi _{g}=0.75$, the switch-up
threshold starts to decrease by transforming the flat segment into a sharp
one. Very small threshold intensities may be an obviously beneficial factor
for applications.

We have also discovered a new shape of the hysteresis curve for the
reflected intensity, as shown in Fig. \ref{figure1}(c) and \ref{figure1}(d).
It is worth to mention that previous studies on the conventional ADCs have
never addressed bistability in reflection \cite{NLG,venugopal}. Our results
predict, on the contrary to the conventional setting (cf. the blue dashed
hysteresis curve), such a novel reflection bistability, wherein one observes the
formation of a loop, pertaining to an unstable segment on the hysteresis
curve, when the value of equal gain is increased to $\chi _{g}=0.25$.  Although the
intensity of the reflected wave increases, it is low compared to the
transmitted intensity [cf. Fig. \ref{figure1}(a)], despite the fact that the
size of the loop (the unstable domain) gets bigger. Remarkable ramifications
are also noticed when the equal gain is set to $\chi _{g}>0.5$. In
particular, one can observe a new form of multistability, transforming from
the looped shape to a novel one, similar to the multistability of the
transmitted intensity. Actually, the amplitude of this novel multistability
is much higher ($\simeq $ by a factor of $10$) than of the multistability of
the transmitted intensity. Above a certain critical value, the development
of the multistability reverts back by forming multiple loops. Also, at $\chi
_{g}\geq 1.5$, lasing behavior is noticed (not shown here). Note that,
although a similar (not exactly the same) reflective bistability has been
observed in semi-conductor amplifiers \cite{maywar} and Fabry-P\'{e}rot
etalons \cite{adam}, reflection multistability has not been reported before
for any feedback structures, including Bragg gratings. Such a unique form of
the reflection bistability may play an essential role in the design
all-optical memory applications, such as AND and NAND gates \cite{adam}. In
addition, the reflection multistability, driven by equal pumping, may open a
new avenue for all-optical signal processing, including memory and switching
in lightwave communication systems.
\begin{figure}[t]
\begin{center}
\includegraphics[width=1\columnwidth]{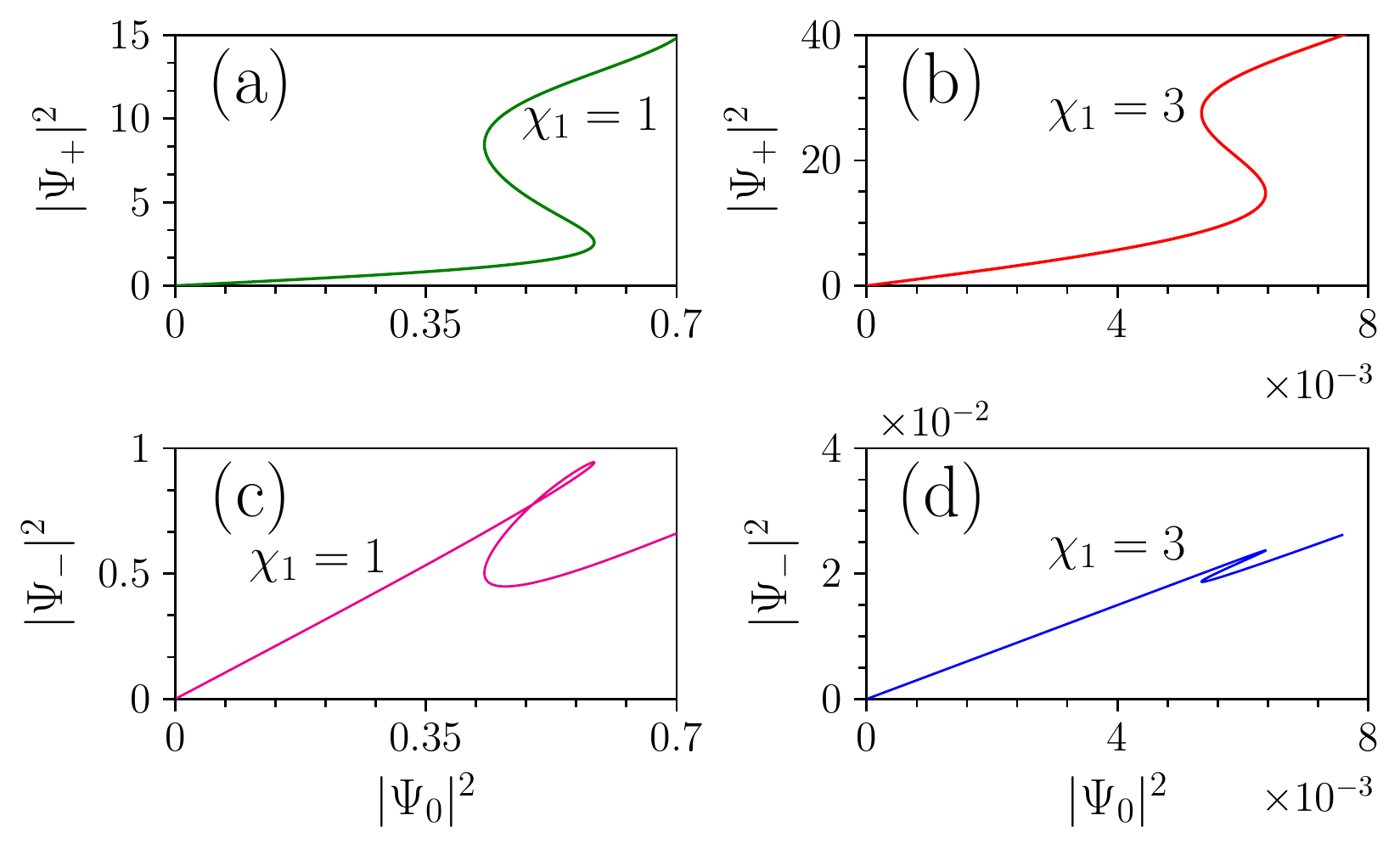}  \includegraphics[width=1%
\columnwidth]{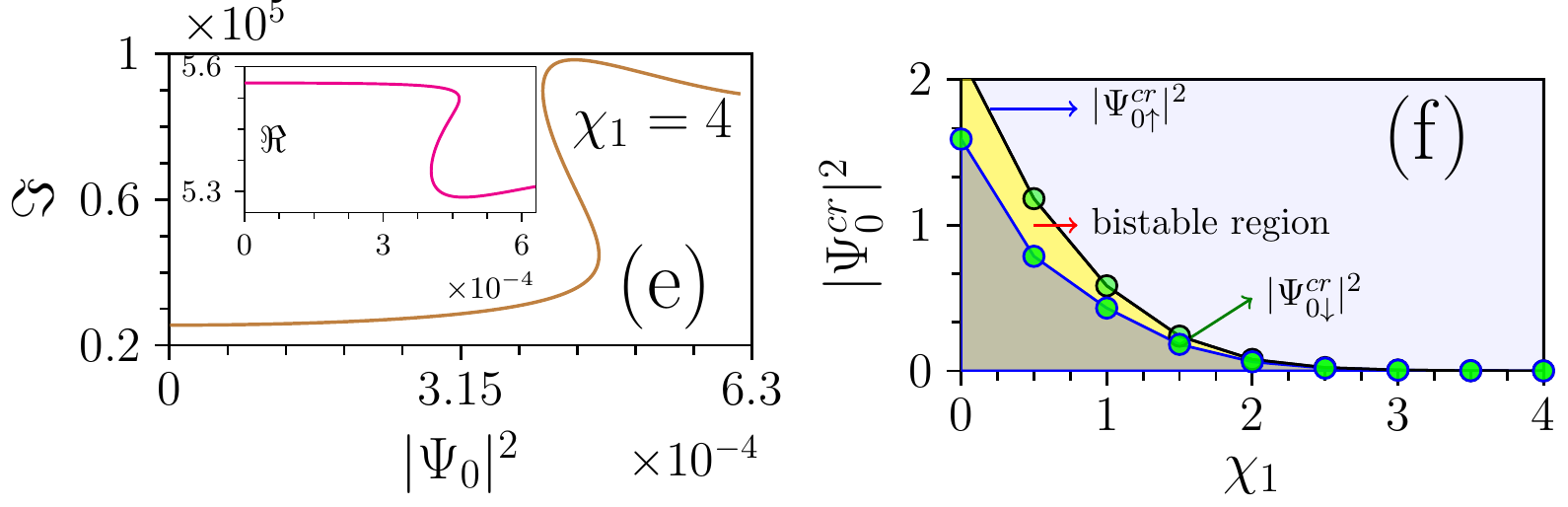}
\end{center}
\caption{The same as in Fig. \protect\ref{figure2}, but for the system with $%
\protect\chi _{1}=2\protect\chi _{2}$ (the gain in the PIM channel twice as
strong as the loss in the NIM one). Other parameters are the same as in Fig.
\protect\ref{figure2}.}
\label{figure3}
\end{figure}

Figure \ref{figure2} summarizes the most essential results of the present
work, by analyzing the role of the balanced gain-loss parameter ($\chi
_{1}=\chi _{2}=\chi $). As was seen in Fig. \ref%
{figure1} and is shown in Fig. \ref{figure2}, upon increasing $\chi $ up to $%
\chi =0.5$, the threshold switch intensity reduces, while featuring huge
amplification of the transmitted intensity in comparison to the nonlinear
ADC without the gain and loss, cf. Ref. \cite{NLG}. Though further increase
of $\chi $ makes the amplification still larger, the hysteresis curve
shrinks in the horizontal direction, gradually losing its bistability, which
implies that the effective feedback is suppressed in the system as the value
of $\chi $ increases, see Fig. \ref{figure2}(f). The role of the reflection
bistability is displayed in Figs. \ref{figure2}(c) -- \ref{figure2}(e),
which demonstrate that the critical intensity for the switch-up bistability
grows with the increase in the value of the equal-gain-loss parameter.
Moreover, the width of the hysteresis curve and the amplitude of the
reflected wave also get reduced, as in the case of the bistability corresponding
to the transmitted intensity. Nonetheless, it is interesting to note that
in both the types of bistabilities, the width of the hysteresis curve remains
unchanged (i.e., the switch-up and down critical intensities of the
reflection and transmission bistabilities remain the same).

To expand the phenomenology, we now fix gain $\chi _{1}$ in the PIM arm to
be twice the value of the loss parameter in the NIM channel, i.e., $\chi
_{1}=2\chi _{2}$, plotting the respective results in Fig. \ref{figure3}.
Unlike the previous case of $\chi _{1}=\chi _{2}$, the bistability is
efficiently sustained even at very high values of the gain and loss
parameters (say, at $\chi _{1}=4,\chi _{2}=2$). In the present case,
switching is possible at ultra-low incident intensity $\sim 10^{-4}$,
whereas the transmissivity jumps to huge values, $\sim 10^{5}$. Such
results, which were not previously reported for nonlinear ADCs, may be
beneficial in the context of all-optical signal processing \cite{radic1994}.
Conversely, the reflective bistability delineated in Figs. \ref{figure3}(c)-%
\ref{figure3}(e) exhibits another novel type of the hysteresis curve, with a
loop different from the one observed in Fig. \ref{figure2}(c). Also, the
first stable state resembles a ramp-like structure extended over a large
range of the input intensity, while the unstable mode forms a loop. It is relevant to note too that, with the increase of the gain and loss coefficients, up to $\chi _{1}=4,\chi _{2}=2$,
in contrast to the transmission bistability, the intensity of the reflected
wave is drastically reduced, following the transition of the hysteresis
curve which loses its looped shape observed at lower values of the gain-loss
parameter, cf. Fig. \ref{figure3}(c).

\begin{figure}[t]
\begin{center}
\includegraphics[width=1\columnwidth]{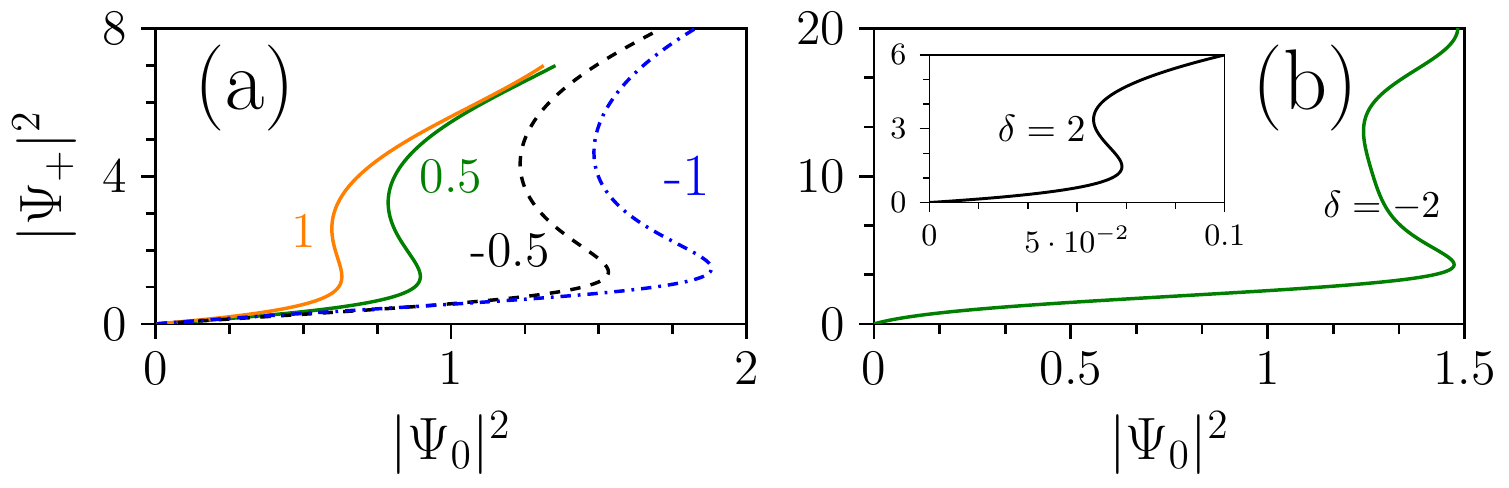} \includegraphics[width=1%
\columnwidth]{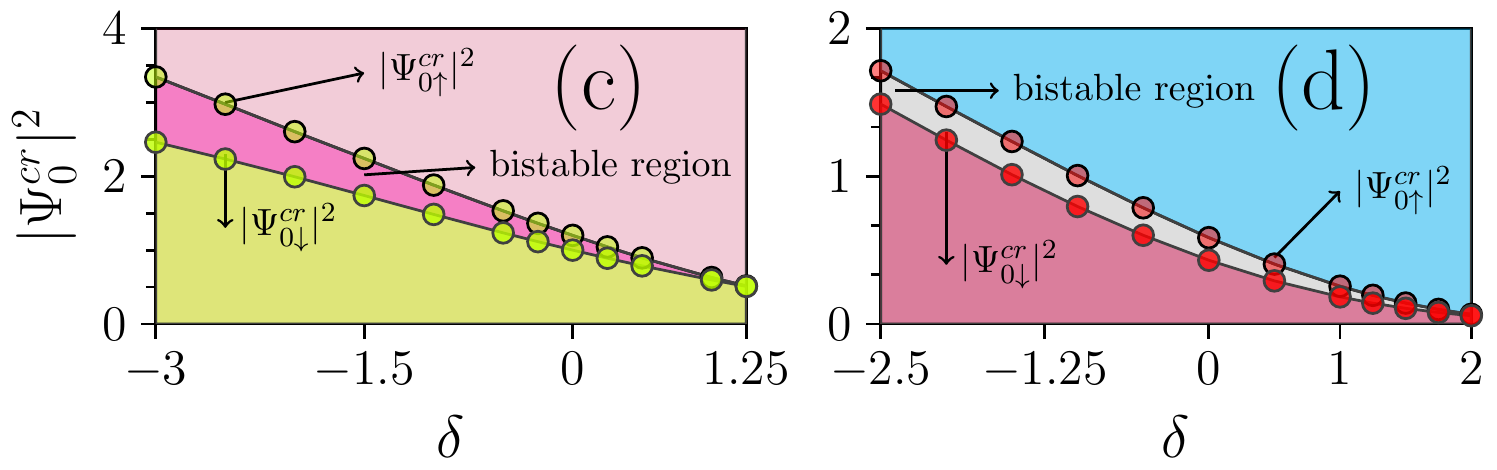}
\end{center}
\caption{The effect of phase-mismatch term, $\protect\delta $, on bistable
states in the systems with $\protect\chi _{1}=\protect\chi _{2}=0.5$ (a) and
$\protect\chi _{1}=2\protect\chi _{2}=1$ (b). Panels (c) and (d) delineate,
respectively, the spectral range of the bistable states for the transmitted
intensity in the systems with $\protect\chi _{1}=\protect\chi _{2}=0.5$ and $%
\protect\chi _{1}=2\protect\chi _{2}=1$. Other parameters are the same as in
Figs. \protect\ref{figure1}.}
\label{figure5}
\end{figure}

We have also studied the effects of the detuning parameter in Eqs. (\ref{eqn:3})
and (\ref{eqn:4}), $\delta $, on the bistability, which was neglected in
previous works \cite{NLG,venugopal,maimistov2012,kazantseva2015,multi2018}, which solely addressed  phase-matched
ADC. Figure \ref{figure5}(a) displays how $\delta $ affects the input-output
curves in the system with equal gain and loss, $\chi _{1}=\chi _{2}$. It is
seen that the increase of $\delta >0$ [for $\delta =0$, one can refer to the
solid orange bistability curve in Fig. \ref{figure2}(a)] tends to suppress
the bistability and hysteresis. On the contrary, $\delta <0$ helps to expand
the hysteresis area, shifting it, as whole, to larger values of the input
intensity.

In addition, Fig. \ref{figure5}(b) shows that $\delta >0$ and $%
\delta <0$ produce qualitatively similar effects in the system with $\chi
_{1}=2\chi _{2}$: comparing to the case of zero mismatch [Fig. \ref{figure3}%
(a)], the negative mismatch shifts the hysteresis to higher values of the
incident intensity, while $\delta >0$ makes it possible to realize the
switching at a very low intensity, while keeping substantial amplification
of the transmitted power. The spectral range of the bistability for the two
types of gain-loss systems is shown in Figs. \ref{figure5}(c) and \ref%
{figure5}(d). In the plot pertaining to equal amount of the gain and loss
[Fig. \ref{figure5}(c)], one observes that the spectral range broadens, in
addition to an increase in the switching (critical) intensities. An opposite
situation is observed for the system in which the gain is twice as strong as
loss, as seen in Fig. \ref{figure5}(d), where the width of the hysteresis
curve (the bistable region) gets reduced for low switching intensities. 

We have also analyzed the role of the ADC's length, concluding that, quite naturally, low values of $L$ suppresses  the hysteresis width, and the bistability tends to 
 disappear, in short couplers. Conversely, an increase of the length 
 helps to expand the hysteresis, in addition to reducing the switching threshold
 (not shown here in detail).
 Thus, one should select an optimum coupler length to design optical bistability 
 with desirable properties.

To summarize, we have reported the role of gain and loss in the operation of
the nonlinear ADC (anti-directional coupler). We have demonstrated that the
gain and loss acting in the positive- and negative-index arms of the coupler
give rise to novel bi- and multi-stability states, admitting dramatic
reduction of the threshold intensity needed for transitions between these
states. These effects can be efficiently controlled by means of the ratio of
the gain and loss coefficients in the two arms, as well as by the phase
mismatch between them.

\textbf{\large Funding.} Science and Engineering Research Board (SERB) of
India: Grants Nos. PDF/2016/002933 and SB/DF/04/2017.


\begin{thebibliography}{99}
\bibitem{bender1} C. M. Bender and S. Boettcher, Phys. Rev. Lett. \textbf{80}%
, 5243-5246 (1998).

\bibitem{dorey} P. Dorey, C. Dunning, and R. Tateo, J. Phys. A: Math. Gen.
\textbf{34}, 5679-5704 (2001).

\bibitem{bender2} C. M. Bender, D. C. Brody, and H. F. Jones, Phys. Rev.
Lett. \textbf{89}, 270401 (2002).

\bibitem{PT1} A. Ruschhaupt, F. Delgado, and J. G. Muga, J. Phys. A:\ Math.
Gen. \textbf{38}, L171-L176 (2005).

\bibitem{PT2} R. El-Ganainy, K. G. Makris, D. N. Christodoulides, and Z. H.
Musslimani, Opt. Lett. \textbf{32}, 2632-2634 (2007).

\bibitem{bender3} C. M. Bender, Rep. Prog. Phys. \textbf{70}, 947-1018
(2007).

\bibitem{review} K. G. Makris, R. El-Ganainy, D. N. Christodoulides, and Z.
H. Musslimani, Int. J. Theor. Phys. \textbf{50}, 1019-1041 (2011).

\bibitem{ptqm} N. Moiseyev, \textit{Non-Hermitian Quantum Mechanics},
(Cambridge University Press, 2011).

\bibitem{Konotop2016RMP} V.~V. Konotop, J.~Yang, and D.~A. Zezyulin,  Rev.
Mod. Phys. \textbf{88}, 035002 (2016).

\bibitem{PTsol2} S. V. Suchkov, A. A. Sukhorukov, J. H. Huang, S. V.
Dmitriev, C. Lee, and Y. S. Kivshar, Laser Photonics Rev. \textbf{10},
177-213 (2016).

\bibitem{el2018non} R.~El-Ganainy, K.~G. Makris, M.~Khajavikhan, Z.~H.
Musslimani, S.~Rotter, and D.~N. Christodoulides, Nature Phys. \textbf{14},
11 (2018).

\bibitem{PhysD} B. A. Malomed, Vortex solitons: Old results and new
perspectives, Physica D, in press;
https://doi.org/10.1016/j.physd.2019.04.009.

\bibitem{Ruter2010Nat} C.~E. R{\"{u}}ter, K.~G. Makris, R.~El-Ganainy, D.~N.
Christodoulides, M.~Segev, and D.~Kip, Nat. Phys. \textbf{6}, 192 (2010).

\bibitem{Kevr-dimer} J. Cuevas, P. G. Kevrekidis, A. Saxena, and A. Khare,
Phys. Rev. A \textbf{88}, 032108 (2013).

\bibitem{Barash-dimer} I. V. Barashenkov, Phys. Rev. A \textbf{90}, 045802
(2014).

\bibitem{Blas} J. D. Huerta Morales, B. M. Rodr\'{\i}gues-Lara, and B. A.
Malomed, Opt. Lett.\textbf{\ 42}, 4402-4405 (2017).

\bibitem{Govind_OL} A.~Govindarajan, A.~K. Sarma, and M.~Lakshmanan,  Opt.
Lett. \textbf{44}, 663 (2019).

\bibitem{Driben} R. Driben and B. A. Malomed, Opt. Lett. \textbf{36},
4323-4325 (2011).

\bibitem{Barash} N. V. Alexeeva, I. V. Barashenkov, A. A. Sukhorukov, and Y.
S. Kivshar, Phys. Rev. A \textbf{85}, 063837 (2012).

\bibitem{Jensen1982CCR} S.~Jensen, IEEE J. Quantum Electron. \textbf{18},
1580 (1982).

\bibitem{stegeman1988bistability} G.~Stegeman, G.~Assanto, R.~Zanoni,
C.~Seaton, E.~Garmire, A.~Maradudin, R.~Reinisch, and G.~Vitrant,  Appl.
Phys. Lett. \textbf{52}, 869 (1988).

\bibitem{Snyder} A. W. Snyder, D. J. Mitchell, L. Poladian, D. R. Rowland,
and Y. Chen, J. Opt. Soc. Am. B \textbf{8}, 2102-2118 (1991).

\bibitem{NL2012} N.~M. Litchinitser, Science \textbf{337}, 1054 (2012).

\bibitem{Cloaking2006Sci} D.~Schurig, J.~Mock, B.~Justice, S.~A. Cummer,
J.~B. Pendry, A.~Starr, and D.~Smith, Science \textbf{314}, 977 (2006).

\bibitem{Engheta2006} N.~Engheta and R.~W. Ziolkowski, \emph{Metamaterials:
physics and engineering explorations} (John Wiley \& Sons 2006).

\bibitem{maimistov2007nonlinear}
A.~I. Maimistov and I.~R. Gabitov, Eur. Phys. J-Spec. Top. \textbf{147}, 265 (2007).

\bibitem{lapine2014}
M.~Lapine, I.~V. Shadrivov, and Y.~S. Kivshar,  Rev. Mod. Phys. \textbf{86}, 1093 (2014).


\bibitem{winfulob} H.~G. Winful, J.~Marburger, and E.~Garmire, Appl. Phys.
Lett. \textbf{35}, 379 (1979).

\bibitem{Cerda} E. A. Cerda-M\'{e}ndez, D. Sarkar, D. N. Krizhanovskii, S.
S. Gavrilov, K. Biermann, M. S. Skolnick, and P. V. Santos, Phys. Rev. Lett.
\textbf{111}, 146401 (2013).

\bibitem{Aceves} A. B. Aceves and S. Wabnitz, Phys. Lett. A \textbf{141},
37-42 (1989).

\bibitem{Christo} D. N. Christodoulides and R. I. Joseph, Phys. Rev. Lett. \textbf{62},
1746-1748 (1989).

\bibitem{deSterke} C. M. de Sterke and J. E. Sipe,  Progr.
Optics \textbf{33}, 203-260 (1994).

\bibitem{Krug} B. J. Eggleton, R. E. Slusher, C. M. de Sterke, P. A. Krug,
and J. E. Sipe, Phys. Rev. Lett. \textbf{76},
1627-1630 (1996).

\bibitem{NLG} N.~M. Litchinitser, I.~R. Gabitov, and A.~I. Maimistov,  Phys.
Rev. Lett. \textbf{99}, 113902 (2007).

\bibitem{leve2007finite} R.~J. LeVeque, \emph{Finite difference methods for
ordinary and partial differential equations: steady-state and time-dependent
problems}, volume~98 (SIAM 2007).

\bibitem{Iooss} G. Iooss and D. D. Joseph, \textit{Elementary Stability and
Bifurcation} \textit{Theory} (Springer: Berlin, 1980).

\bibitem{radic1994} S.~Radic, N.~George, and G.~P. Agrawal, Opt. Lett.
\textbf{19}, 1789 (1994).

\bibitem{chen1987gap} W.~Chen and D.~Mills, Phys. Rev. Lett. \textbf{58},
160 (1987).

\bibitem{winful1991mi} H.~G. Winful, R.~Zamir, and S.~Feldman, Appl. Phys.
Lett. \textbf{58}, 1001 (1991).

\bibitem{d2004bright} G.~D'Aguanno, N.~Mattiucci, M.~Scalora, and M.~J.
Bloemer, Phys. Rev. Lett. \textbf{93}, 213902 (2004).

\bibitem{venugopal} G.~Venugopal, Z.~Kudyshev, and N.~M. Litchinitser,  IEEE
J. Sel. Top. Quantum Electron. \textbf{18}, 753 (2012).

\bibitem{maimistov2012}
A.~Maimistov and E.~Kazantseva, Optics and Spectroscopy \textbf{112}, 264 (2012).

\bibitem{kazantseva2015}
E.~Kazantseva and A.~Maimistov, Optics
and Spectroscopy \textbf{119}, 776 (2015).

\bibitem{multi2018}
K.~Nithyanandan, A.~S. Ali, K.~Porsezian, M.~Nishad, P.~T. Dinda, and P.~Grelu, Opt.
Commun. \textbf{416}, 145 (2018).


\bibitem{maywar} D. N. Maywar and G. P. Agrawal, IEEE J. Quantum Electron.
\textbf{34}, 2364 (1998).

\bibitem{adam} M. J. Adams, Opt. Quant. Electron. \textbf{19}, S37 (1987).

\end{thebibliography}
\end{document}